\documentclass[graybox]{svmult}

\usepackage{mathptmx}       
\usepackage{helvet}         
\usepackage{courier}        
\usepackage{type1cm}        
\usepackage{makeidx}         
\usepackage{graphicx}        
\usepackage{multicol}        
\usepackage[bottom]{footmisc}

\makeindex             

\begin{document}

\title*{Non-parametric latent modeling and network clustering}
\titlerunning{Non-parametric latent modeling and network clustering} 
\author{Fran\c{c}ois Bavaud}
\authorrunning{Non-parametric latent modeling and network clustering} 
\institute{Fran\c{c}ois Bavaud \at University of Lausanne, \email{fbavaud@unil.ch}}
%
%
\maketitle


\abstract{The paper exposes a non-parametric approach to latent and co-latent modeling of bivariate data, based upon alternating minimization of the Kullback-Leibler divergence (EM algorithm) for complete log-linear models. For categorical data, the iterative algorithm generates a soft clustering of both rows and columns of the contingency table. Well-known results are systematically revisited, and some variants are presumably original. In particular, the consideration of square contingency tables induces a clustering algorithm for weighted networks, differing from spectral clustering or modularity maximization techniques. Also, we present a co-clustering algorithm applicable to HMM models of general kind,  distinct from the Baum-Welch algorithm. 
Three case studies illustrate the theory.}



\section{Introduction: parametric and non-parametric mixtures}
Two variables can be dependent, yet conditionally independent given a third one, that is $X\perp Y|G$ but $X\not\perp Y$: in bivariate {\em latent models} of dependence $M$, joint bivariate probabilities $P(x,y)$ express as
\begin{equation}
\label{found}
P(x,y)=\sum_{g=1}^m p(x,y,g)=\sum_{g=1}^m p(g)p(x|g)p(y|g)
\end{equation} 
where $x$, $y$, $g$ denote the values of $X$, $Y$, $G$, and $p(x,y,g)$ their joint probability. 

\

Bivariate data, such as summarized by normalized contingency tables $F(x,y)=\frac{n(x,y)}{n(\bullet,\bullet)}$, where $n(x,y)$ counts the number of individuals in $x\in X$ and $y\in Y$, can be approached by 
latent modeling,  consisting in 
inferring a suitable model $P(x,y)\in M$ of the form (\ref{found}), typically 
closest to the  observed  frequencies $F(x,y)$   in the maximum-likelihood sense, or in the least squares sense.  Mixture (\ref{found}) also defines {\em memberships} $p(g|x)=p(x|g)p(g)/p(x)$ and $p(g|y)$; hence latent modeling also performs {\em model-based clustering}, 
 assigning  observations $x$ and $y$ among groups $g=1,\ldots,m$. 

 \
 
 Latent modeling and clustering count among the most active data-analytic research trends of the last decades. The literature is simply too enormous to cite even a few valuable contributions, often (re-)discovered independently among workers in various application fields. Most approaches are parametric, typically defining  $p(x|g)$ and $p(y|g)$ as exponential distributions of some kind, such as the multivariate normal (continuous case) or the multinomial (discrete case) (see e.g. Govaert and   Nadif 2013 and references therein). Parametric modelling allows further 
hyperparametric Bayesian processing, as in latent Dirichlet allocation (Blei et al. 2003). 
 
 \

By contrast, we focus on non-parametric models specified by the whole family of log-linear {\em complete models}  ${\cal M}$ corresponding to $X\perp Y|G$, namely (see e.g. Christensen 2006)\begin{displaymath}
{\cal M}=\{ p\: |\: \ln p(x,y,g)=a(x,g)+b(y,g)+c\}
\end{displaymath}
Equivalently, 
\begin{displaymath}
{\cal M}=\{ p\: |\: p(x,y,g)=\frac{p(x,\bullet,g)\: p(\bullet, y,g)}{p(\bullet,\bullet,g)}\}
\end{displaymath}
where ``$\bullet$" denotes the summation over the replaced argument. 
The corresponding class of bivariate models $M$  of the form (\ref{found}) simply reads 
$M=\{ P\: |\: P(x,y)=\sum_g p(x,y,g)\equiv p(x,y,\bullet)\: \: \mbox{, \small for some } p\in{\cal M}\}$.

\begin{figure}[b]
\sidecaption
\includegraphics[scale=.13]{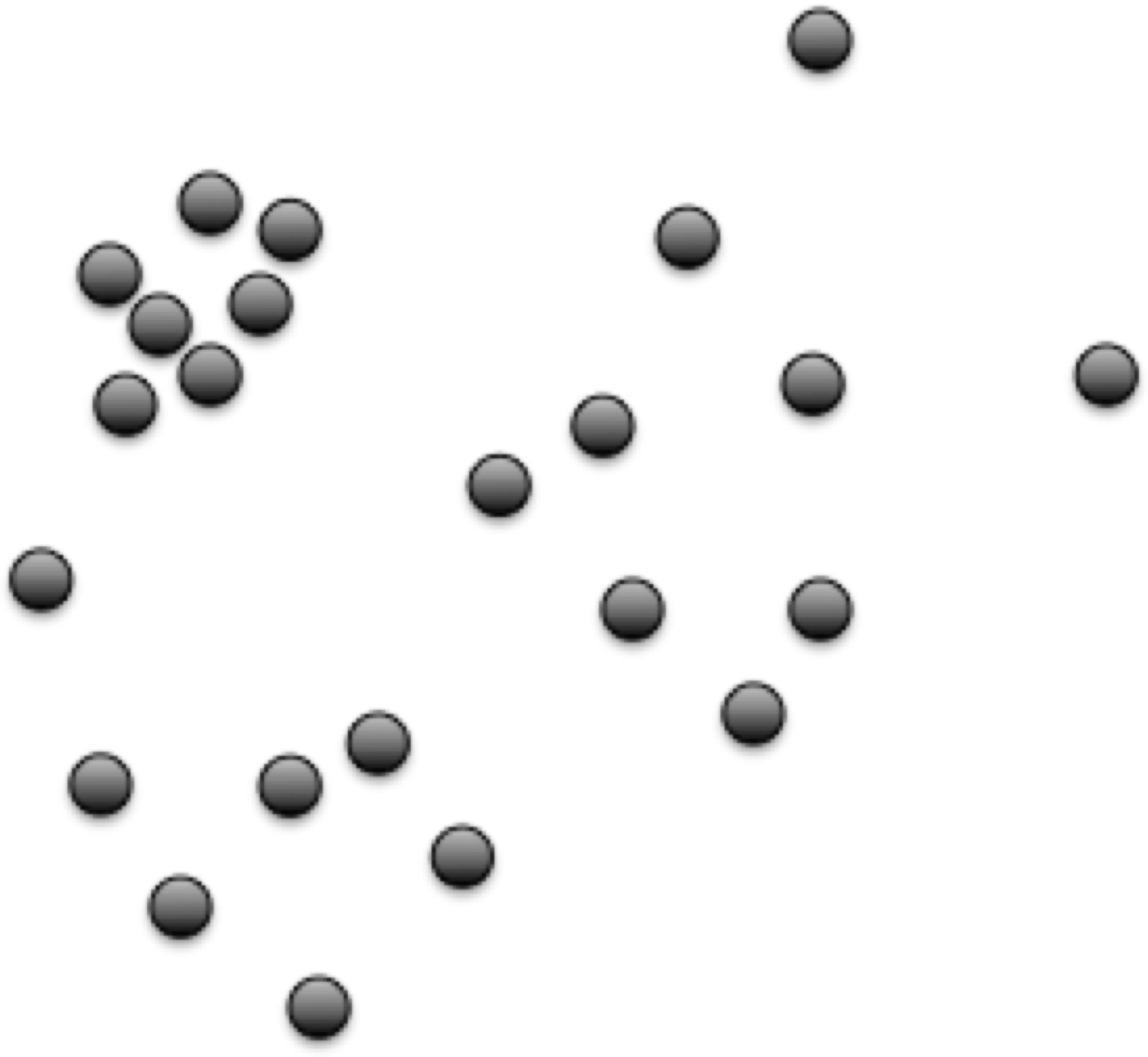}\hspace{-0.25cm}
\includegraphics[scale=.13]{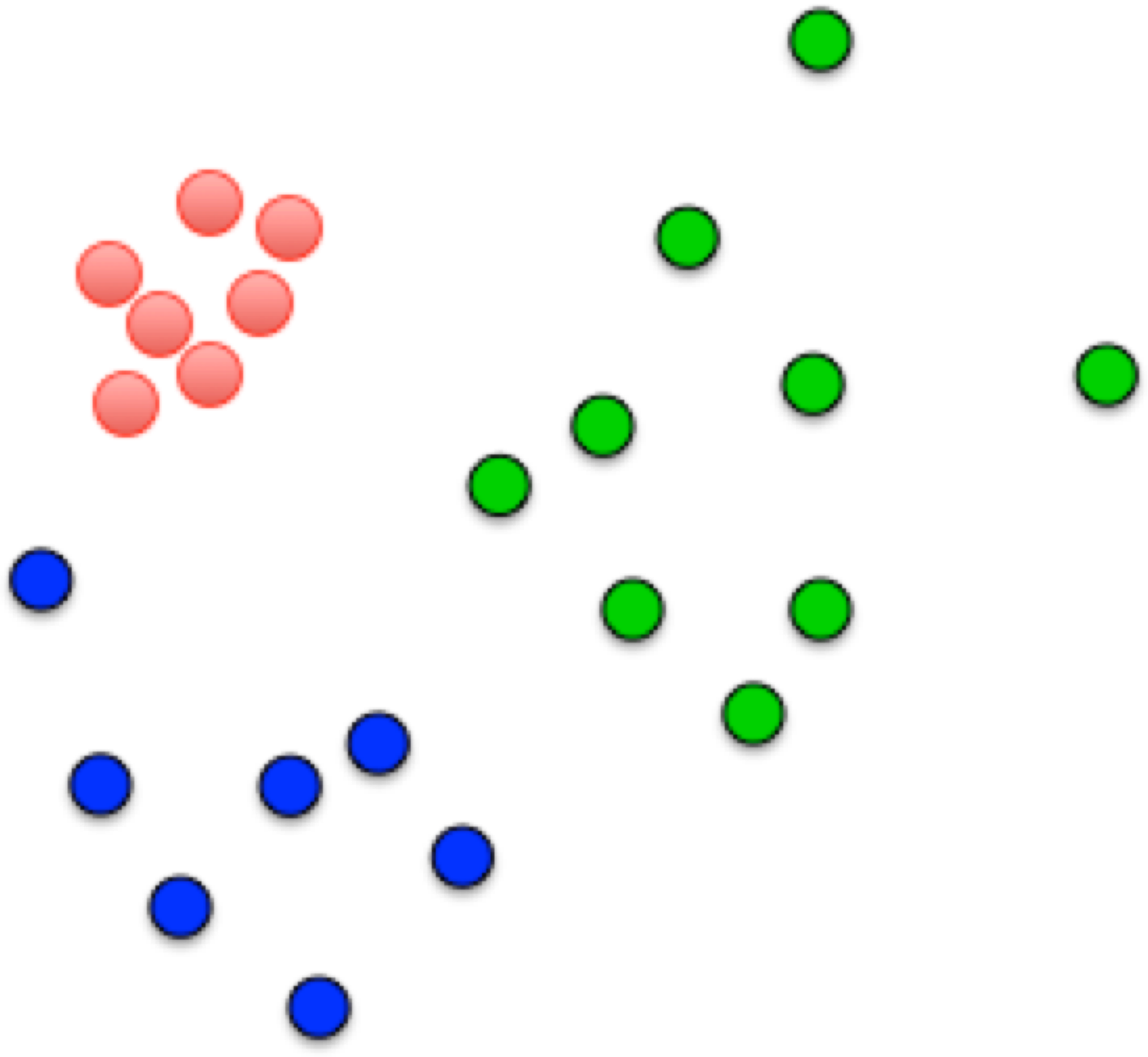}
%
%
\caption{Left: observed data, where $(x,y)$ are the object coordinates. Right: complete data $(x,y,g)$, where the group $g$ is labeled by a color.
In psychological terms, $(x,y)$ is the {\em stimulus}, and $(x,y,g)$ the {\em percept}, emphasizing  the  EM-algorithm as a possible model for cognition.}
\label{complete_data}       
\end{figure}
 
 \

Observations consist of the joint empirical distribution $F(x,y)$, normalized to $F(\bullet,\bullet)=1$. In latent modeling, one can think of the observer as a color-blind agent perceiving only the margin $f(x,y,\bullet)$ of the complete distribution $f(x,y,g)$, but not the color (or group) $g$  itself (see Fig. \ref{complete_data}). Initially, any member $f$ of the set 
\begin{displaymath}
{\cal D}=\{f\: |\: f(x,y,\bullet)=F(x,y)\}
\end{displaymath}
seems equally compatible with the observations $F$, and the role of a clustering algorithm precisely consists in selecting a few  good  candidates $f\in {\cal D}$, or even a unique one,  bringing color to the observer.

\

This paper exposes a non-parametric approach to latent and co-latent modeling of bivariate data, based upon alternating minimization of the Kullback-Leibler divergence (EM algorithm) for complete log-linear models (section \ref{secEM}). For categorical data, the iterative algorithm generates a soft clustering of both rows and columns of the contingency table. Well-known results are systematically revisited, and some variants are presumably original. In particular, the consideration of square contingency tables induces a clustering algorithm for weighted networks, differing from spectral clustering or modularity maximization techniques (section \ref{netwclu56}). Also, we present a co-clustering algorithm applicable to HMM models of general kind,  distinct from the Baum-Welch algorithm. 
Three case studies illustrate the theory: latent (co-)betrayed
clustering of a term-document matrix (section \ref{reuttdm}), latent clustering of spatial flows (section \ref{lbbhsmsm}), and 
latent co-clustering of bigrams in French  (section \ref{lbbh}).

\section{EM latent clustering: a concise derivation from first principles}
\label{secEM}
The {\em alternating minimisation} procedure
(Csisz\'ar and Tusn\'ady 1984) provides an arguably elegant derivation of the EM algorithm; see also e.g.  Cover and Thomas (1991) or  Bavaud (2009). The maximum likelihood model $\hat{P}\in M$ of the form (\ref{found}) minimizes the Kullback-Leibler divergence $K()$
\begin{displaymath}
\hat{P}=\arg\min_{P\in M} \: K(F\| P)
\qquad\qquad K(F\| P)=\sum_{x,y}F(x,y)\ln\frac{F(x,y)}{P(x,y)}
\end{displaymath}
where $F(x,y)$ denotes the empirical bivariate distribution. On the other hand, the complete 
Kullback-Leibler divergence $K(f\| p)=\sum_{x,y,g}f(x,y,g)\: \ln\frac{f(x,y,g)}{p(x,y,g)}$, where 
$f(x,y,g)$ is the empirical ``complete" distribution (see fig. \ref{complete_data}), enjoys the following properties (see e.g. Bavaud (2009) for the proofs, standard in Information Theory): 
\begin{equation}
\label{mstep}
\hat{p}(x,y,g):=\arg\min_{p\in {\cal M}}\: K(f||p)=\frac{f(x,\bullet, g)\: f(\bullet, y,g)}{f(\bullet,\bullet, g)}
\qquad\qquad \mbox{\bf M-step}
\end{equation}
\begin{equation}
\label{estep}
\tilde{f}(x,y,g):=\arg\min_{f\in {\cal D}} \: K(f||p)=\frac{p(x,y, g)}{p(x,y,\bullet)}\: F(x,y)
\qquad\qquad\quad \mbox{\bf E-step}
\end{equation}
Furthermore, $\min_{f\in {\cal D}}\: K(f||p)=K(F||P)$, and thus
\begin{displaymath}
\min_{P\in M} \: K(F||P)=\min_{p\in {\cal M}}\min_{f\in {\cal D}}\:  K(f||p)
\end{displaymath}
Hence, starting from some complete model $p^{(0)}\in {\cal M}$, the EM-sequence $f^{(t+1)}:=\tilde{f}[p^{(t)}]$ defined in (\ref{estep}) and 
$p^{(t+1)}:=\hat{p}[f^{(t+1)}]$ defined in  (\ref{mstep}) converges towards a {\em local minimum} of $K(f||p)$. Observe  the margins to coincide after a single EM-cycle  in the sense $p^{(t)}(x,\bullet,\bullet)=F(x,\bullet)$ and
$p^{(t)}(\bullet,y,\bullet)=F(\bullet,y)$ for all $t\ge1$. 

\

For completeness sake, note that ${\cal D}$  and ${\cal M}$ are {\em closed} in the following sense, as they are in other instances of  the EM algorithm in general. Critically and crucially: 
\begin{enumerate}
  \item[i)] ${\cal D}$ is convex, that is closed under additive mixtures $\lambda f_1+(1-\lambda)f_2$; this turns out to be the case for maximum entropy problems in general. 
 \item[ii)] ${\cal M}$ is log-convex, that is closed under multiplicative mixtures $ p_1^\lambda p_2^{(1-\lambda)}/Z(\lambda)$ where $Z(\lambda)$ is a normalization constant; this is the case for exponential models, as well as for non-parametric log-linear models in general. 
 \end{enumerate}

\subsection{Latent co-clustering}
\label{COsecEM}
Co-clustering describes the situation where each of the observed variables is attached to a {\em distinct} latent variable, the latter being mutually associated. That is, $X\perp Y|(U,V)$, $X\perp V|U$ and 
$Y\perp U|V$ while $X\not\perp Y$, and  $U\not\perp V$ in general. Equivalently, $X\to U\to V\to Y$ form a ``Markov chain", in the sense of Cover and Thomas (1991). 
Bivariate joint probabilities express as 
\begin{equation}
\label{cofound}
P(x,y)=\sum_{u=1}^{m_1}\sum_{v=1}^{m_2} p(x,y,u,v)=\sum_{u,v} p(u,v)p(x|u)p(y|v)
\end{equation} 
Complete models ${\cal M}$, restricted models $M$ and complete empirical distributions ${\cal D}$ are 
\begin{eqnarray}
{\cal M} & = & \{ p\: |\: p(x,y,u,v)=\frac{p(x\bullet u\bullet)\: p(\bullet y\bullet v)\: p(\bullet  \bullet u v)}{p(\bullet\bullet u\bullet)\: p(\bullet\bullet\bullet v)} \} \\
M & = & \{P\: |\: P(x,y)=p(x,y,\bullet,\bullet)\: \: \mbox{with}\: \: p\in {\cal M}\} 
\\
{\cal D} & = & \{f\: |\: f(x,y,\bullet,\bullet)=F(x,y)\} 
\end{eqnarray}
where $F(x,y)$ denotes the observed empirical distribution. The steps of the former section apply again, yielding the EM algorithm 
\begin{equation}
\label{comstep}
\hat{p}(x,y,u,v):=\arg\min_{p\in {\cal M}}\:  K(f||p)=\frac{f(x\bullet u\bullet)\: f(\bullet y\bullet v)\: f(\bullet  \bullet u v)}{f(\bullet\bullet u\bullet)\: f(\bullet\bullet\bullet v)}
\qquad \mbox{\bf M-step}
\end{equation}
\begin{equation}
\label{coestep}
\tilde{f}(x,y,u,v):=\arg\min_{f\in {\cal D}} \: K(f||p)=\frac{p(x,y, u,v)}{p(x,y,\bullet,\bullet)}\: F(x,y)
\qquad\qquad\quad \mbox{\bf E-step}
\end{equation}
where $K(f\| p)=\sum_{x,y,u,v}f(x,y,u,v)\: \ln\frac{f(x,y,u.v)}{p(x,y,u,v)}$ measures the divergence of  the complete observations from the complete model. 

\subsection{Matrix and tensor algebra for contingency tables}
The material of sections (\ref{secEM}) and (\ref{COsecEM}) holds {\em irrespectively of the continuous or discrete  nature of $X$ and $Y$}: in the continous case, integrals simply replace sums. In the  discrete setting, addressed here, categories are numbered as 
 $i=1,\ldots,n$ for $X$, as $k=1,\ldots,p$ for $Y$ and as $g=1,\ldots,m$ for $G$. Data consist  of the relative $n\times p$ contingency table $F_{ik}$ normalized to $F_{\bullet\bullet}=1$. 
 
 \subsubsection{Latent co-clustering}
 Co-clustering models and complete models express as 
 \begin{equation}
\label{cofund}
P_{ik}=\sum_{u=1}^{m_1}\sum_{v=1}^{m_2}c_{uv}\: a_i^u\: b_k^v
\qquad\qquad p_{ikuv}=c_{uv}\: a_i^u\: b_k^v
\end{equation}
\begin{enumerate}
   \item[$\bullet$] where   $c_{uv}=P(U=u,V=v)=p(\bullet\bullet uv)$, obeying $c_{\bullet\bullet}=1$, is the {\em joint  latent distribution} of row, respectively column groups $u$ and $v$
  \item[$\bullet$]  $a_i^u=p(i\bullet u\bullet)/p(\bullet \bullet u \bullet)$ (with $a_{\bullet}^u=1$) is the {\em row distribution conditionally to the row group $U=u$}, also referred to as {\em emission probability} (section \ref{netwclu56})
  \item[$\bullet$] $b_k^v=p(\bullet k \bullet v)/
 p(\bullet \bullet\bullet v)$ (with $b_{\bullet}^v=1$) is the column distribution or emission probability conditionally to the column group $V=v$.
\end{enumerate}
  Hence, a  complete model $p$ is entirely determined by the triple $(C,A,B)$, where $C=(c_{uv})$ is $m_1\times m_2$ and normalized to unity, $A=(a_i^u)$ is $n\times m_1$ and $B=(b_k^v)$ is $p\times m_2$, both row-standardized.

 \
 
It is straightforward to show that the successive application of the E-step (\ref{coestep}) and the M-step (\ref{comstep}) to  $p\equiv (C,A,B)$ yields the new complete model $\ddot{p}\equiv (\ddot{C},\ddot{A},\ddot{B})$ with 
\begin{eqnarray}
\ddot{c}_{uv} & = & c_{uv}\: \sum_{jl}\frac{F_{jl}}{P_{jl}}\: a_j^u\: b_l^v \label{core1}\\
\ddot{a}_i^u & = & a_i^u \: \frac{\sum_{lv'}c_{uv'}\: \frac{F_{il}}{P_{il}}\: b_l^{v'}}
{\sum_{jlv'}c_{uv'}\frac{F_{jl}}{P_{jl}}\:  a_j^u\:  b_l^{v'}} \label{core2}\\
\ddot{b}_k^v & = & b_k^v \: \frac{\sum_{ju'}c_{u'v}\: \frac{F_{jk}}{P_{jk}}\: a_j^{u'}}
{\sum_{jlu'}c_{u'v}\frac{F_{jl}}{P_{jl}}\:  a_j^{u'}\: b_l^{v}} \label{core3}
\end{eqnarray}
 Also, after a single EM cycle, margins are respected, that is $\ddot{P}_{i\bullet}=F_{i\bullet}$ and
 $\ddot{P}_{\bullet k}=F_{ \bullet k}$.  
 
 \
 
In hard clustering, rows $i$ are attached to a single   group denoted $u[i]$, that is $a_i^u=0$ unless $u=u[i]$; similarly, $b_k^v=0$ unless $v=v[k]$. Restricting $P$ in (\ref{cofund}) to 
hard clustering yields  {\em block clustering}, for which $K(F||P)=I(X:Y)-I(U:V)$, where $I()$ is the mutual information (e.g. Kullback (1959); Bavaud  (2000); Dhillon et al. (2003)). 
  
 \
 
 The set $M$ of models $P$ of the form  (\ref{cofund}) is convex, with extreme points consisting of hard clusterings. $K(F\| P)$ being convex in $P$, its minimum is attained for convex mixtures of hard clusterings, that is for {\em soft clusterings}.

 \subsubsection{Latent clustering}
 Setting $m_1=m_2=m$ and $C$ diagonal with $c_{gh}=\rho_g\: \delta_{gh}$ yields the latent model 
  \begin{equation}
\label{fundmat}
P_{ik}=\sum_{g=1}^{m}\rho_g \: a_i^g\: b_k^g
\qquad\qquad p_{ikg}=\rho_g \: a_i^g\: b_k^g
\end{equation}
 together with the corresponding EM-iteration  $p\equiv (\rho,A,B)\to  \ddot{p}\equiv (\ddot{\rho},\ddot{A},\ddot{B})$, namely
 \begin{equation}
\label{hjkktuzt}
\ddot{\rho}_{g}=\rho_g\: \kappa_g
\qquad\qquad
\ddot{a}_i^g=a_i^g \frac{\sum_l b_l^g\frac{F_{il}}{P_{il}}}{\kappa_g}
\qquad\qquad
\ddot{b}_k^g=b_k^g \frac{\sum_j a_j^g\frac{F_{jk}}{P_{jk}}}{\kappa_g}
\end{equation}
where $\kappa_g=\sum_{jl}a_j^gb_l^g\frac{F_{jl}}{P_{jl}}$. Similar, if not equivalent updating rules have been proposed in information retrieval and natural language processing (Saul and  Pereira 1997; Hofmann 1999), as well as in the {\em non-negative matrix factorization} framework (Lee and Seung 2001; 
 Finesso and Spreij 2006). 

\

By construction, families of latent models (\ref{fundmat}) $M_m$ with $m$ groups are nested in the sense $M_m\subseteq M_{m+1}$.   

The case $m=1$ amounts to {\em independence models} $P_{ik}=a_i b_k$, for which the fixed point $\ddot{a}_i=F_{i\bullet}$ and $\ddot{b}_k=F_{\bullet k}$ is, as expected,  reached after a single iteration, irrespectively of the initial values of $a$ and $b$. 

By contrast, $m\ge \mbox{rank}(F)$ generates {\em saturated models}, exactly reproducing the observed contingency table. For instance, assume that $m=p=\mbox{rank}(F)\le n$; then taking
$a_{i}^g=F_{ig}/F_{\bullet g}$, $b_k^g=\delta_{kg}$ and $\rho_g=F_{\bullet g}$ (which already constitutes a fixed point of (\ref{hjkktuzt})) evidently satisfies $P_{ik}=F_{ik}$.

\subsection{Case study I: Reuters 21578 term-document matrix}
\label{reuttdm}
\begin{figure}
\begin{center}
\includegraphics[width=3.8cm]{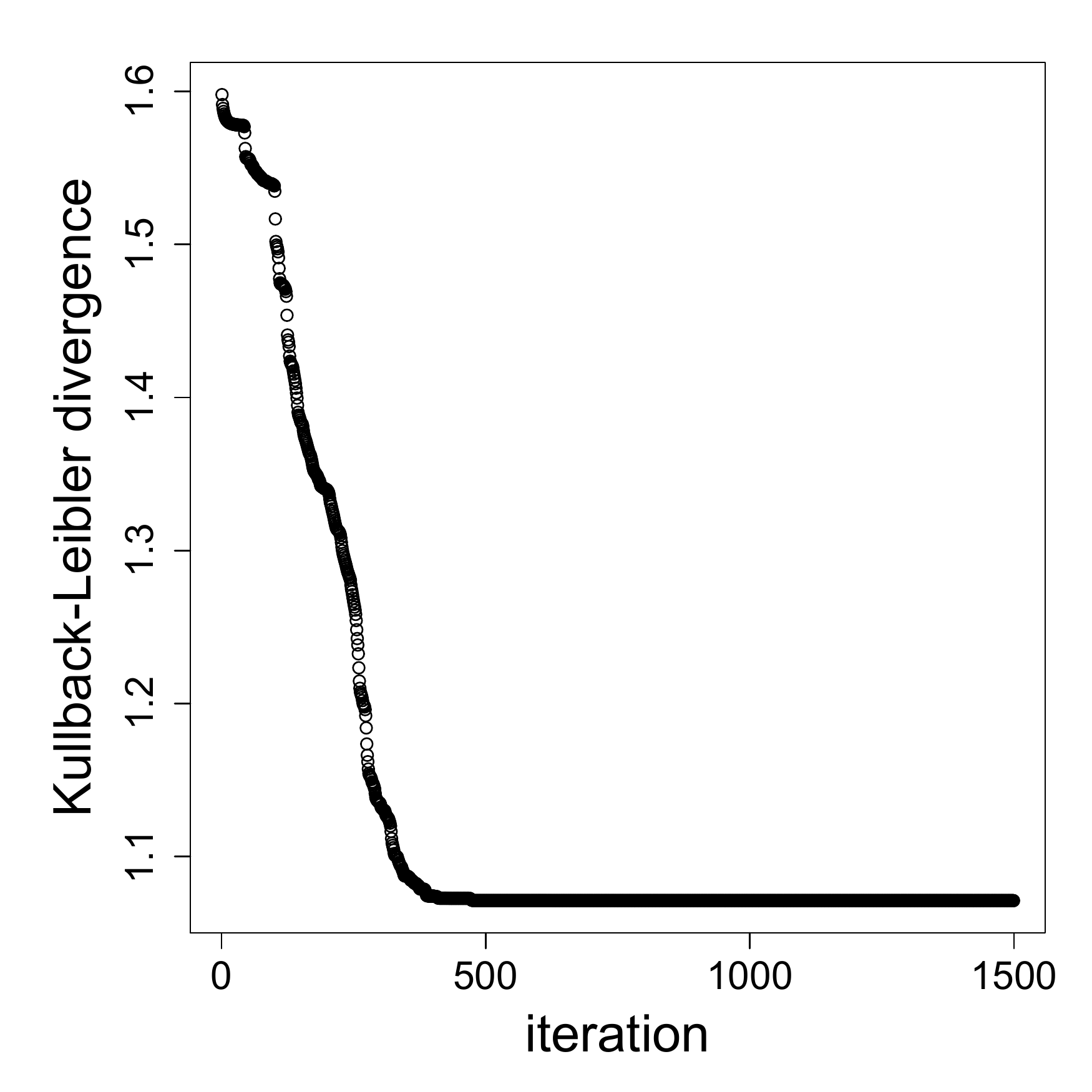}
\includegraphics[width=3.8cm]{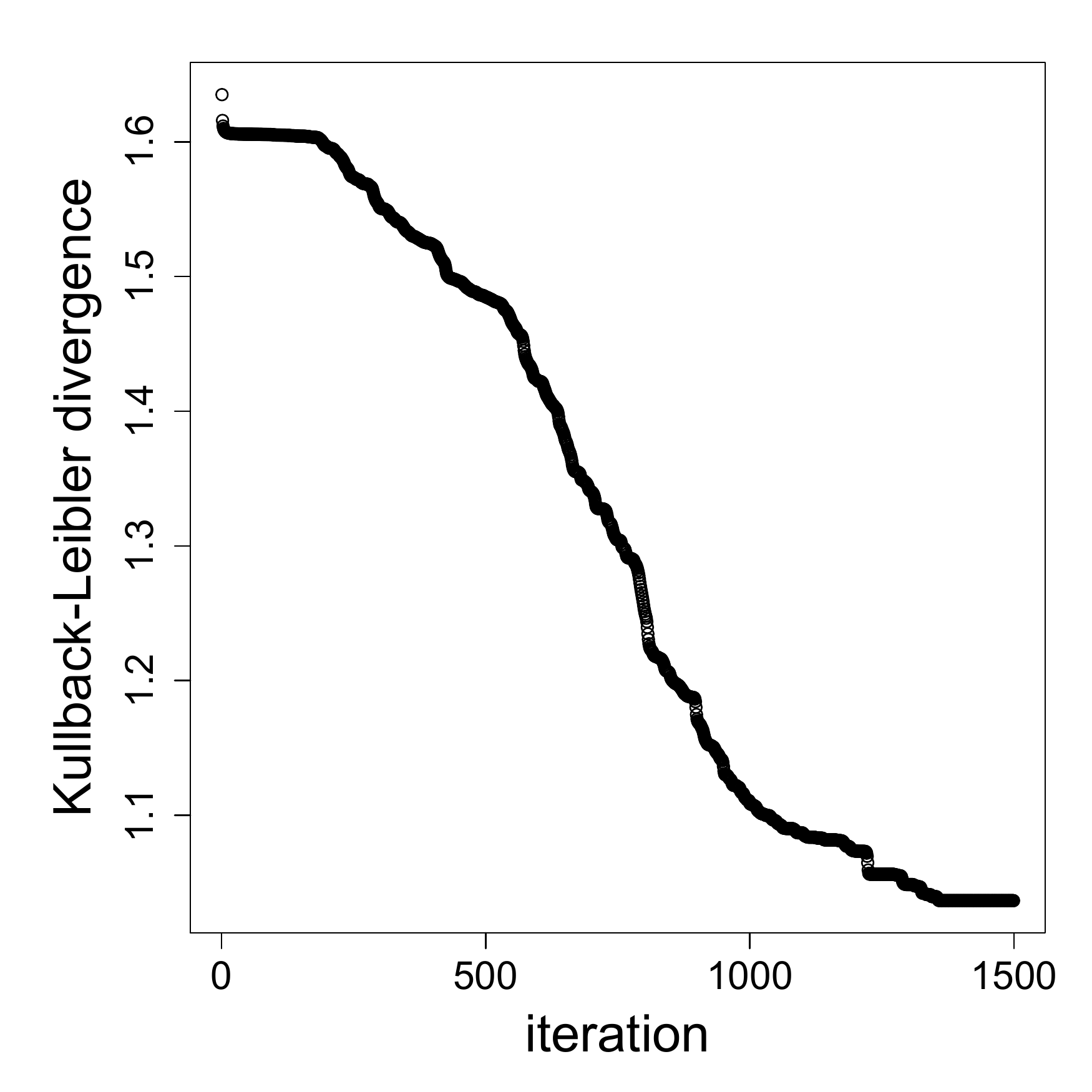}
\includegraphics[width=3.8cm]{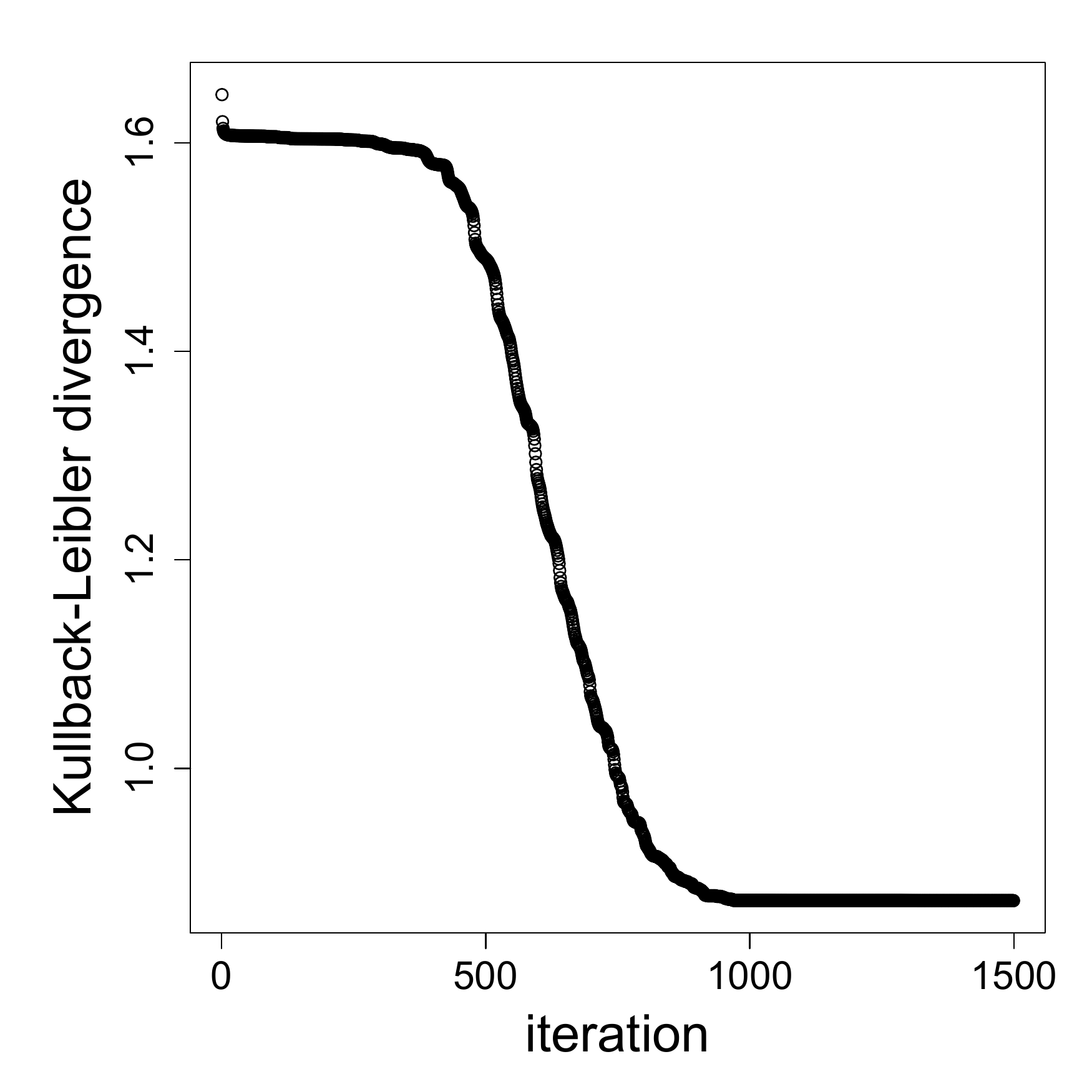}
\caption{Case study I: convergence of the latent and co-latent iterating procedure. Left: latent model with $m=3$. Middle: co-latent model with $(m_1,m_2)=(3,4)$. Right: co-latent model with $(m_1,m_2)=(4,4)$.}
\label{KLreuters}
\end{center}
\end{figure}

The   $n\times p= 20\times 1266$ document-term normalized matrix $F$, constituting the {\em Reuters 21578} dataset, is accessible through the R package {\bf  tm} (Feinerer et al.  2008). The co-clustering algorithm (\ref{core1}) (\ref{core2}) (\ref{core3}) is started by randomly assigning uniformly each document  to a single row group $u=1,\ldots,m_1$, and by uniformly assigning  each term  to a single column group $v=1,\ldots,m_2$. The procedure turns out  to converge after about 1000 iterations (figure \ref{KLreuters}), yielding a locally minimal value $K_{m_1m_2}$  of the Kullback-Leibler divergence. By construction, $K_{m_1m_2}$ decreases with $m_1$ and $m_2$. Latent clustering (\ref{hjkktuzt}) with $m$ groups is performed analogously, yielding a locally minimal value $K_m$. 

\

Experiments with three or four groups yield the typical results $K_3=1.071180$ $>$ $K_{33}= 1.058654$
 $>$ 
$K_{43}=1.038837$
 $>$ 
$K_{34}=1.036647$
 $>$ 
$K_4=0.877754$
 $>$ 
$K_{44}=0.873071$. The above ordering is expected, although inversions are frequently observed, under differing random initial configurations. Model selection procedures, not addressed here, should naturally consider in addition the degrees of freedom, larger for co-clustering models. The latter do not appear as particularly rewarding here (at least for the experiments performed, and in contrast to the results associated to case study III of section \ref{lbbh}): indeed, joint latent distributions $C$ turn out to be ``maximally sparse", meaning that  row groups  $u$ and  column groups  $v$ are essentially the same. Finally, each of the 20 documents of the Reuters 2157 dataset happens to belong to a single row group (hard clusters), while only a minority of the 1266 terms (say about 20\%) belong to two or more  column groups (soft clusters).

 \vspace{-1cm}

\section{Network clustering}
\label{netwclu56}
When the observed categories $x$ and $y$ belong to the same set  indexed by $i,j=1,\ldots,n$, the relative square contingency table $F_{ij}$ defines a {\em directed weighted network} on $n$ vertices: $F_{ij}$ is the weight of edge $(ij)$, $F_{i\bullet}$ is the outweight of vertex $i$ (relative outdegree) and 
$F_{\bullet i}$ its inweight  $i$ (relative indegree), all normalized to unity. Frequently, $F_{ij}$ counts  the relative number of units initially at vertex $i$, and at vertex $j$  after some fixed time. Examples abound in spatial migration, spatial commuting, social mobility, opinion shifts, confusion matrices, textual dynamics, etc.

\

A further restriction, natural in many applications of latent network modeling, consists in identifying the row and column emission probabilities, that is in requiring $b_i^g=a_i^g$. This condition   generates four families of nested latent network models of increasing flexibility, namely
     \begin{eqnarray}
P_{ij}& = & \sum_{g=1}^m  \rho_g \: a_i^g\: a_j^g   \hspace{3.35cm}
\mbox{latent (symmetric) network model}\label{cas1}\\
P_{ij}& = & \sum_{u,v=1}^m c_{uv} \: a_i^u\: a_j^v\quad\mbox{with $c_{uv}=c_{vu}$} 
\qquad \mbox{co-latent symmetric network model} \label{cas2a}  \\
P_{ij}& = & \sum_{u,v=1}^m c_{uv} \: a_i^u\: a_j^v\quad\mbox{with $c_{u\bullet}=c_{\bullet u}$} 
\qquad \mbox{co-latent MH network model} \label{cas2b}  \\
P_{ij}& = & \sum_{u,v=1}^m c_{uv} \: a_i^u\: a_j^v\hspace{3.1cm}
\mbox{co-latent general network model}  \hspace{1cm}
\label{cas33}  
\end{eqnarray}

Models (\ref{cas1}) and (\ref{cas2a}) $P=P'$, making latent and co-latent {\em symmetric} clustering suitable for {\em unoriented weighted networks} with $F_{ij}=F_{ji}$. By contrast, unrestricted co-latent models (\ref{cas33})  describes general {\em oriented weighted networks}. 
Symmetric matrices $F=(F_{ij})$ appear naturally in reversible random walks on networks, or in spatial modeling where they measure the spatial interaction between regions (spatial weights), and constitute a weighted version of the adjacency matrix, referred to as an {\em exchange matrix} by the author (Bavaud 2014 and references therein; see also Berger and Snell 1957). 

\

Latent models (\ref{cas1}) are positive semi-definite or {\em diffusive}, that is endowed with {\em non-negative eigenvalues}, characteristic of a continuous propagation process  from a one place to its neighbours. In particular, the diagonal part of $P$ in  (\ref{cas1}) cannot be too small. In contrast, co-latent symmetrical network models (\ref{cas2a}) are flexible enough to describe  phenomena such as  bipartition or periodic alternation, implying negative eigenvalues. 

 \

The condition (\ref{cas2b})  of {\em marginal homogeneity} (MH) on the joint latent distribution $C$  is inherited by the restricted models, in the sense $P_{i\bullet}=P_{\bullet i}$. They constitute appropriate models for the bigram distributions of single categorical sequences (of length $N$, constituted of $n$ types), for which $F_{i\bullet}=F_{\bullet i}+O(N^{-1})$; see the case study III of Section \ref{lbbh}. Formulation (\ref{cas2b}) describes $m$ hidden states related by a Markov transition matrix $p(v|u)=c_{uv}/c_{u\bullet}$, as well as $n$ observed states related to the hidden ones by the {\em emission probabilities} $a_i^u=p(i|u)$. Noticeably enough, (\ref{cas2b}) precisely encompasses the ingredients of the {\em hidden Markov models} (HMM) (see e.g. Rabiner 1989).

 \subsection{Network latent clustering}
 \label{kljhlnnnn99}
Approximating $F$ by $P$ in (\ref{cas1}) amounts in performing a {\em soft network clustering}: the {\em membership} of vertex $i$ in group $g$ (of weight $\rho_g$) is
\begin{displaymath}
z_{ig}=p(i|g)=\frac{p(i)p(g|i)}{p(g)}=\frac{f_i\:  a_i^g}{\rho_g}
\qquad\mbox{with}\quad f_i=F_{i\bullet}=F_{\bullet i}\quad\mbox{and}\quad \rho_g =\sum_{i=1}^n f_i \: z_{ig}\enspace. 
\end{displaymath}
EM-updating rules for   memberships (instead of  emission probabilities, for a change)
\begin{equation}
\label{epsydotb}
P_{ij}=f_if_j \sum_{g=1}^m\frac{z_{ig}z_{jg}}{\rho_g}
\qquad\qquad \ddot{z}_{ig}=z_{ig} 
\: \sum_j \frac{F_{ij}}{P_{ij}}\frac{f_jz_{jg}}{\rho_g}
\qquad\qquad \ddot{\rho}_g=\sum_i f_i \:  \ddot{z}_{ig}
\end{equation} 
 define a soft clustering iterative algorithm for unoriented weighted networks, presumably   original.

  \subsection{Case study II: inter-cantonal Swiss migrations} 
\label{lbbhsmsm}
Consider the  $n\times n$ matrix  $N=(N_{ij})$ of inter-cantonal migratory flows in Switzerland, counting the number of people inhabiting canton 
$i$ in 1980 and canton $j$ in 1985, $i,j=1,\ldots, n=26$, for a total of $ \mbox{sum}(N)=
6'039'313$ inhabitants, $93\%$ of which lie on the diagonal (stayers).  The symmetric, normalized matrix $F=\frac12(N+N')/N_{\bullet\bullet}$ is diffusive, largely dominated by its diagonal. As a consequence, direct application of algorithm (\ref{epsydotb}) from an initial random cantons-to-groups assignation produces somewhat erratic results: a matrix $F=(F_{ij})$ too close to the identity matrix $I=(\delta_{ij})$ cannot  by reasonably approximated by the latent model (\ref{cas1}), unless $m=n$, where each canton belongs to its own group. 

Here, the difficulty lies in the shortness of the observation period (5 years, smaller than the average moving time), making the off-diagonal contribution $1-\mbox{trace}(F)$ too small. Multiplying the observation period by a factor $\lambda>1$ generates, up to $O(\lambda^2)$, a modified relative flow $\tilde{F}_{ij}=\lambda F_{ij}+(1-\lambda)\delta_{ij}f_i$, where $f_i=F_{i\bullet}=F_{\bullet i}$ is the weight of canton $i$. The modified
$\tilde{F}$ is normalized, symmetric, possesses unchanged vertex weights $\tilde{F}_{i\bullet}=f_i$, and its off-diagonal contribution is  multiplied by $\lambda$. Of course, $\lambda$ cannot be too large, in order to insure the non-negativity of $\tilde{F}$ ($\lambda\le 6.9$ here) as well as its semi-positive definiteness ($\lambda\le 6.4$ here). 

\begin{figure}[h]
\sidecaption
\includegraphics[scale=.2]{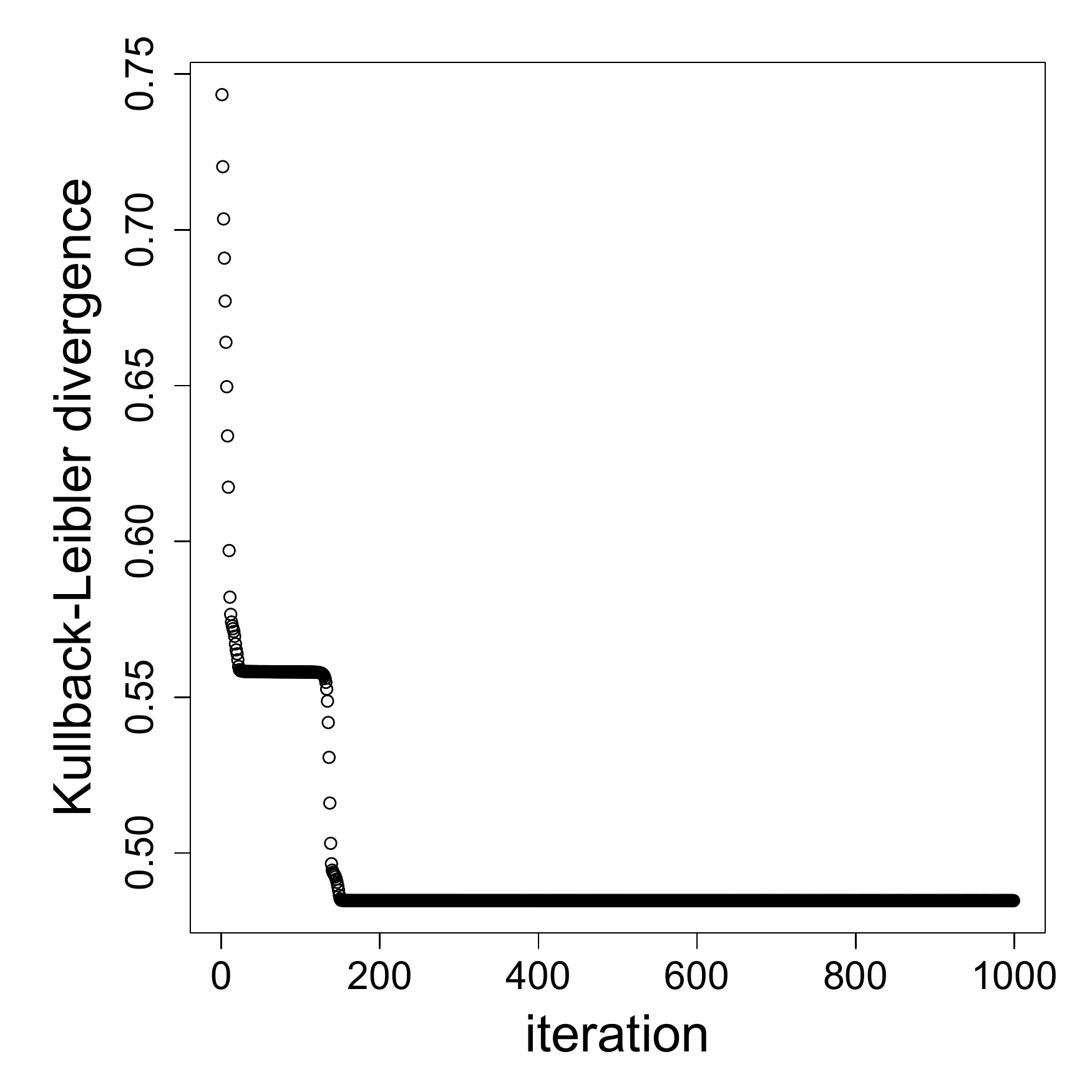}\hspace{0.35cm}
\includegraphics[scale=.2]{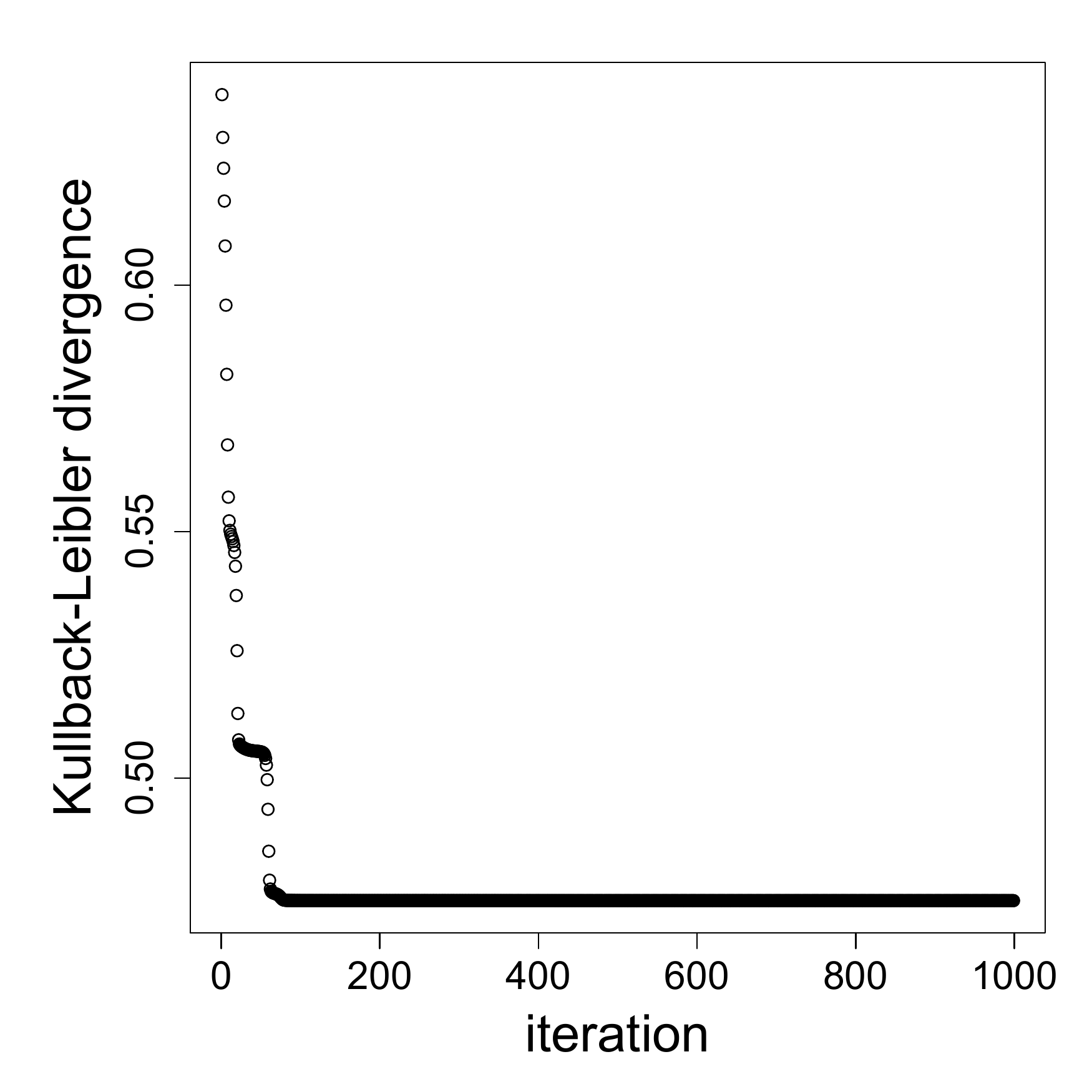}
%
%
\caption{Decrease of the Kullback-Leibler divergence for the two realizations of figure \ref{MAPswiss}, respectively. Horizontal plateaux correspond to metastable minima in the learning of the latent structure, followed by the rapid discovery of a better fit.}
\label{KLswiss}        
\end{figure}

\begin{figure}[h]
\hspace{-0.2cm}\includegraphics[scale=.2]{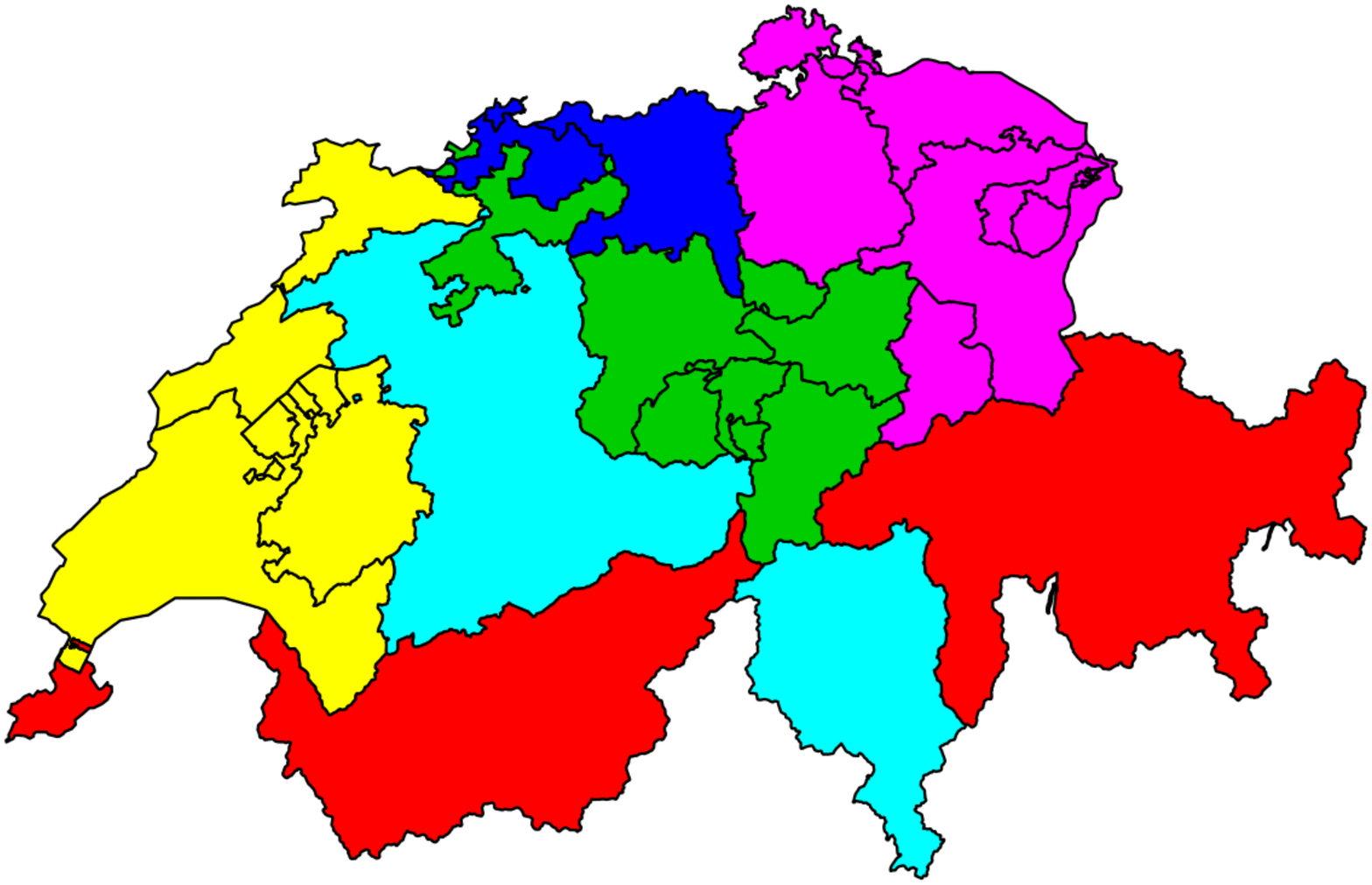}\hspace{0.35cm}
\includegraphics[scale=.2]{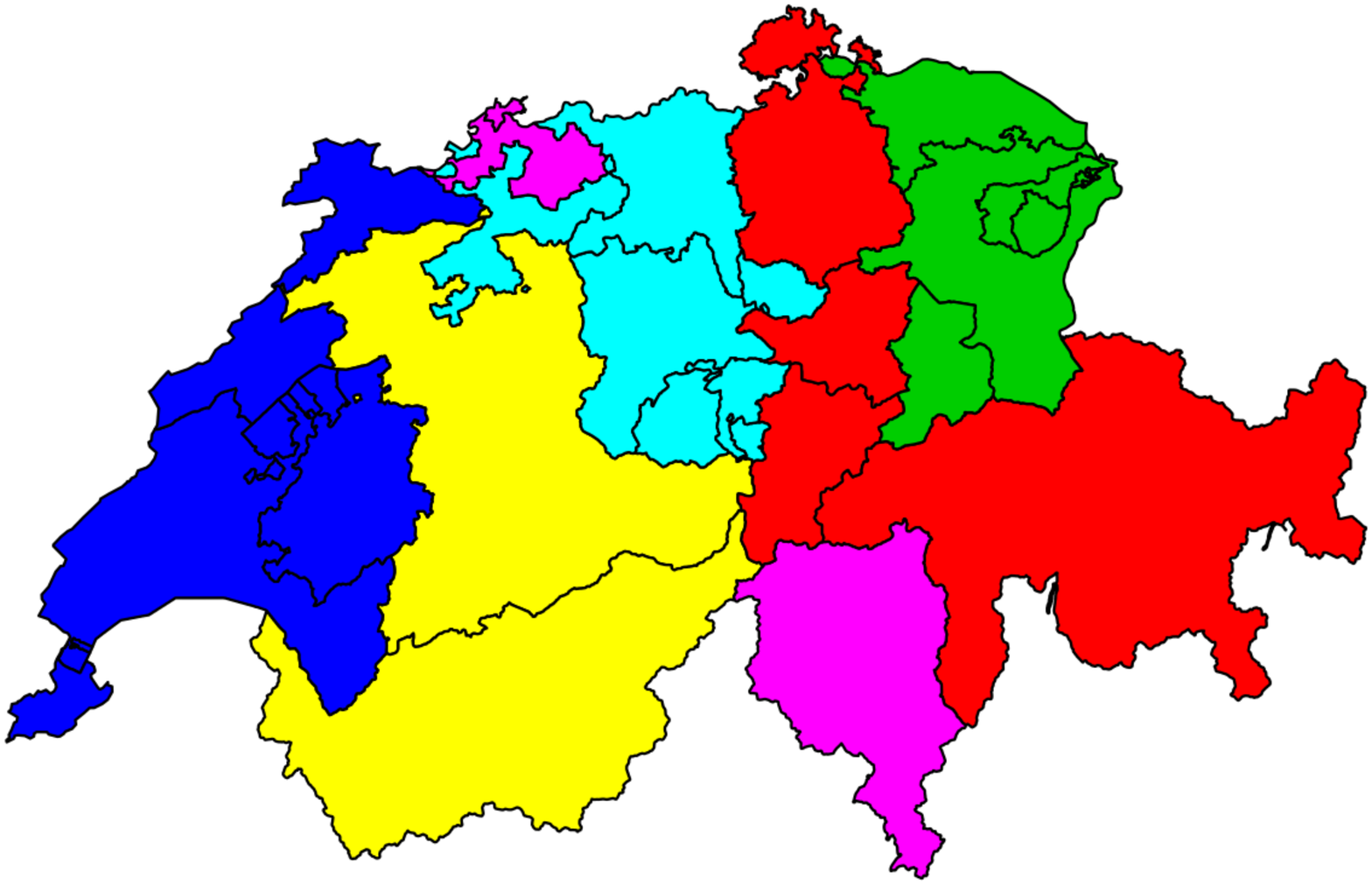}
%
%
\vspace*{-0.6cm}
\caption{Case study II:  two realizations of the network latent clustering algorithm (\ref{epsydotb}), applied to the modified flow matrix $\tilde{F}$, with random initial assignment to $m=6$ groups, and final hard assignment  of canton $i$ to group $\arg\max_g z_{ig}$.}
\label{MAPswiss}        
\end{figure}

Typical realizations of (\ref{epsydotb}), with $\lambda=5$, are depicted in figures (\ref{KLswiss}) and 
(\ref{MAPswiss}): as expected, spatially close regions tend to be regrouped.

\vspace*{-0.8cm}
\subsection{Network general co-clustering}
\label{kjgkg1199}
Latent co-clustering (\ref{cas33}) applies to  contingency tables $F$ of general kind, possibly asymmetric or marginally inhomogeneous, and possibly exhibiting diffusivity, alternation, or a mixture of them. 
Implementing the common emission probabilities constraint in the M-step (\ref{comstep}) yields together with (\ref{cas33})  the updating rule
\begin{equation}
\label{ncchmm}
\ddot{c}_{uv}=c_{uv}\: \sum_{ij}\frac{F_{ij}}{P_{ij}}\: a_i^u\: a_j^v\qquad\qquad
\ddot{a}_{i}^u=a_i^u\: \frac{\sum_{j'v'}(c_{uv'}\frac{F_{ij'}}{P_{ij'}}+c_{v'u}\frac{F_{j'i}}{P_{j'i}})\: a_{j'}^{v'}}{\sum_{i'j'v'}(c_{uv'}\frac{F_{i'j'}}{P_{i'j'}}+c_{v'u}\frac{F_{j'i'}}{P_{j'i'}})\: a_{i'}^{u}\: a_{j'}^{v'}}
\end{equation}
Let us recall that the classical Baum-Welch  algorithm handles the HMM modeling of a single (large) sequence of tokens, with (almost) marginally homogeneous bigram counts. By contrast, the presumably original iterative algorithm
(\ref{ncchmm}) also seems  to be able to handle  marginally inhomogeneous network data, such that aggregates of smaller sequences. Further experimentations are needed  at this stage to gauge  the generality of model (\ref{cas33}), and the efficiency of  algorithm (\ref{ncchmm}).

\

For symmetric data $F=F'$, the symmetric model (\ref{cas2a}) can be tackled by  (\ref{ncchmm}) above, with the simplifying circumstance that the additive symmetrizing occurring in the  numerator and denominator of $\ddot{a}_{i}^u$ is not needed anymore,  provided that the initial joint probability $c_{uv}$ is symmetrical, which automatically insures the symmetry of 
further iterates $\ddot{c}_{uv}$. 

\vspace*{0cm}

\subsection{Case study III: modeling bigrams} 
\label{lbbh}
\vspace*{-1cm}
\begin{table}
\begin{center}\begin{tabular}{|c|rrrr|}\hline
\multicolumn{1}{|c}{group}   & 1& 2 &  3 & 4 \\ \hline
\_ &     &    &  61  &    \\
a  &     &    &     &  25  \\
b  &  3   &    &    &   \\
c  &  10   &   1 &  1  &    \\
d  &   12  &  2  &    &    \\
e  &  1   &    &    & 45 \\
f & 4    &    &    &    \\
g &  3   & 1   &    &    \\
h &  2   &    &    &  1  \\
i &  1   & 3   & 13   &  5  \\
j &  2   &    &    &    \\
k &     &    &    &    \\
l &  17   &    &    &  4  \\
m &   8  &    &  2  &    \\
n &  17   & 9   &    &    \\
o &     &    &    & 13   \\
p &  8   &    &  1  &    \\
q &     &  5  &    &    \\
r &   10  &  6  &  4  &  3  \\
s &  6   &  28  &    &    \\
t &    7 &  26  &    &    \\
u &     & 8   &  10  &   3 \\
v &   7  &    &    &    \\
w &     &    &    &    \\
x &     &  2  &    &    \\
y &  1   &    &    &     \\
z &     &  1  &    & \\ \hline
 \end{tabular}
\hspace*{2cm} \begin{tabular}{|c|rrrr|}\hline
\multicolumn{1}{|c}{group}   & 1& 2 &  3 & 4 \\ \hline
\_ &     &    &  100  &    \\
a  &     &    &     &  100  \\
b  & 89    &    &  11  &   \\
c  &  87   &   5 &  8  &    \\
d  &  90   &  10  &    &    \\
e  &  2   &    & 1   &  97 \\
f &  87   &    &    &  13  \\
g &  71   &   26  &    & 3   \\
h &  65   &    &    &  35  \\
i &   4  &  9  &   61 &   26 \\
j &   100  &    &    &    \\
k &     &  100  &    &    \\
l &   77  &   2 &    &  21  \\
m &  76   &    & 24   &    \\
n &     & 53   &  47  &    \\
o &     &    &    &  100  \\
p &  89   &    &  9   &  2  \\
q &     &  88  &    & 12   \\
r &  40   & 21   &  22 &  17  \\
s &  20   & 80   &    &    \\
t &  26   &  74  &    &    \\
u &     &  27  &  57  &  16  \\
v &  100   &    &    &    \\
w &   100  &    &    &    \\
x &     &  100  &    &    \\
y &  60   &  40  &    &     \\
z &     &  100  &    & \\ \hline
 \end{tabular} 
\hspace*{2cm}  \begin{tabular}{|c|rrrr|}\hline
  & 1& 2 &  3 & 4 \\ \hline
1 &  0   &   0 &  1  &  21 \\  
2 &  0   &  3  &  12  & 2 \\  
3 &  17   &  7  &  0   &  7 \\  
4 & 5   &  8  &  17  &  0 \\ \hline
 \multicolumn{5}{c}{} \\
 \multicolumn{5}{c}{} \\
 \multicolumn{5}{c}{} \\
 \multicolumn{5}{c}{} \\
 \multicolumn{5}{c}{} \\
 \multicolumn{5}{c}{} \\ \hline
& 1& 2 &  3 & 4 \\ \hline
1 &  0   &   0 &  7  &  93 \\  
2 &  0   &  15  &  71  & 14 \\  
3 &  54   &  24  &  0   &  22 \\  
4 & 17   &  27  &  57  &  0 \\ \hline
 \multicolumn{5}{c}{} \\
 \multicolumn{5}{c}{} \\
 \multicolumn{5}{c}{} \\
 \multicolumn{5}{c}{} \\
 \multicolumn{5}{c}{} \\
 \multicolumn{5}{c}{} \\
\multicolumn{2}{r|}{1} & \multicolumn{3}{l}{22} \\ 
\multicolumn{2}{r|}{2} & \multicolumn{3}{l}{18} \\ 
\multicolumn{2}{r|}{3} & \multicolumn{3}{l}{31} \\ 
\multicolumn{2}{r|}{4} & \multicolumn{3}{l}{30} \\ 
 \end{tabular} 
\caption{Case study III: emission probabilities $A$ (left), memberships $Z$ (middle), joint latent distribution $C$ (right, top), latent probability transition matrix $W$ (right, middle) and its corresponding stationary distribution $\pi$ (right, bottom). All values are multiplied by 100 and rounded to the nearest integer.}
\label{jkjkjkjkjkjkjk}
\end{center}
\end{table}

 \vspace{-0.5cm}
We consider the first chapters of the French novel ``La B\^{e}te humaine" by Emile Zola (1890). After suppressing all punctuation, accents and separators with exception of the blank space, and converting
 upper-case letters to lower-case, we are left with a sequence of $N=725'000$ tokens on $n=27$ types (the alphabet + the space), containing 724'999 pairs of successive tokens or {\em bigrams}. The resulting $n\times n $ normalized contingency table $F=(F_{ij})$ is far from symmetric (for instance, the bigram {\tt\bf  qu} occurs 6'707 times, while {\tt\bf  uq} occurs only 23 times), but almost {\em marginally homogenous}, that is $F_{i\bullet}\cong F_{\bullet i}+0(N^{-1})$ (and exactly marginally homogenous if one  starts and finishes the textual sequence with the same type, such as  a blank space). 
 
 Symmetrizing $F$ as $F^s=(F+F')/2$ does not makes it diffusive, and hence unsuitable by latent modelling (\ref{cas1}), because of the importance of large negative eigenvalues in 
 $F^s$, betraying {\em alternation}, typical in linguistic data - think in particular of the vowels-consonants alternation (e.g. Goldsmith and Xanthos 2009). This being said, symmetric co-clustering of $F^s$ (\ref{cas2a}) remains a possible option. 
 
 Table \ref{jkjkjkjkjkjkjk} results from the general co-clustering algorithm  (\ref{ncchmm}) applied on the original, asymmetric  bigram counts $F$ itself. Group 4 mainly emits the vowels, group 3 the blank, group 2 the {\tt s} and {\tt t}, and group 1 other consonants. Alternation is betrayed by the null diagonal of the Markov transition matrix $W$ - with the exception of group 2.
 
The property of marginal homogeneity $F_{i\bullet}= F_{\bullet i}$ permits  in addition to obtain the {\em memberships $Z$ from the emissions $A$}, by first  determining the solution $\rho$ of $\sum_g \rho_g\: a_{i}^g=f_i$, where $f_i=F_{i\bullet}= F_{\bullet i}$ is the relative frequency of letter $i$, and then by defining $z_{ig}=\rho_g\: a_{i}^g/f_i$.

\end{document}